\documentclass[%
 article,showpacs,showkeys,
 amsmath,amssymb,
 reprint,%
floatfix,]{revtex4-1}

\usepackage{graphicx}
\usepackage{dcolumn}
\usepackage{bm}

\usepackage[utf8]{inputenc}
\usepackage[T1]{fontenc}
\usepackage{mathptmx}
\usepackage{etoolbox}
\usepackage{physics}
\usepackage{color}
\usepackage[dvipsnames]{xcolor}
\usepackage{accents}
\usepackage{hyperref}

\newcommand{\ubar}[1]{\underaccent{\bar}{#1}}

\makeatletter
\def\@email#1#2{%
 \endgroup
 \patchcmd{\titleblock@produce}
  {\frontmatter@RRAPformat}
  {\frontmatter@RRAPformat{\produce@RRAP{*#1\href{mailto:#2}{#2}}}\frontmatter@RRAPformat}
  {}{}
}%
\makeatother
\begin{document}
\title{Lattice Boltzmann simulations of Plasma Wakefield Acceleration}
\author{G. Parise$^*$}
\affiliation{Department of Physics, University of Rome "Tor Vergata", Via della Ricerca Scientifica 1, 00133, Rome (Italy)}
\affiliation{INFN, Laboratori Nazionali di Frascati, Via Enrico Fermi 54, 00044, Frascati (Italy)}
\email{gianmarco.parise@lnf.infn.it}
\author{A. Cianchi}
\affiliation{Department of Physics \& INFN, University of Rome "Tor Vergata", Via della Ricerca Scientifica 1, 00133, Rome (Italy)}
\author{A. Del Dotto}%
\affiliation{ENEA, C.R. Brasimone, 40032, Camugnano, Bologna (Italy)}
\author{F. Guglietta}
\affiliation{Helmholtz Institute Erlangen-Nürnberg for Renewable Energy (IEK-11), Forschungszentrum Jülich GmbH, Cauerstraße 1, 91058, Erlangen (Germany)}%
\author{A. R. Rossi}
\affiliation{INFN, Section of Milan, via Celoria, 16, 20133, Milan (Italy)}
 \author{M. Sbragaglia}
 \affiliation{Department of Physics \& INFN, University of Rome "Tor Vergata", Via della Ricerca Scientifica 1, 00133, Rome (Italy) }

\begin{abstract}
We explore a novel simulation route for {\it Plasma Wakefield Acceleration} (PWFA) by using the computational method known as the {\it Lattice Boltzmann Method} (LBM). LBM is based on a discretization of the continuum kinetic theory while assuring the convergence towards hydrodynamics for coarse-grained fields (i.e., density, velocity, etc.). LBM is an established numerical analysis tool in computational fluid dynamics, able to efficiently bridge between kinetic theory and hydrodynamics, but its application in the context of PWFA has never been investigated so far. This paper aims at filling this gap. Results of LBM simulations for PWFA are discussed and compared with those of a code (Architect) implementing a \emph{Cold Fluid} (CF) model for the plasma. In the hydrodynamic framework, we discuss the importance of regularization effects related to diffusion properties intrinsic of the LBM, allowing to go beyond the traditional CF approximations. Issues on computational efficiency are also addressed. 
\end{abstract}
\maketitle
\section{Introduction}\label{sec:intro}
 Plasma acceleration is one of the most interesting technique to develop compact and cost affordable particle accelerators. 
 A perturbation in a neutral plasma induced by a driver pulse~\cite{litos2014high, esarey2009physics,blumenfeld2007energy} produces a strong electric filed, called \textit{wakefield}, due to the separation between positive and negative charges.  
 This wakefield ($1-100~$GV/m) is several orders of magnitude larger than the one generated by the conventional radio frequency machines ($\lesssim 50$~MV/m). 
 With this intense field it is possible to accelerate particles at high energy with smaller and cheaper accelerating structures promoting their usage in small labs, industries, hospitals and universities~\cite{doyle2019future}.
 Particle wakefield acceleration (PWFA)~\cite{chen1985acceleration} relies on the use of a particle bunch as a driver, creating the wave due to Coulomb repulsion; laser wakefield acceleration (LWFA)~\cite{esarey2009physics} makes use of a high-intense laser ($>10^{18}~\mbox{W/cm}^2$) and the laser ponderomotive force generates the wave.\\
Plasma acceleration is a complex, multi-scale process: numerical simulations are indispensable to well design an experiment and to completely understand it. Indeed, since a lot of variables are involved in plasma acceleration experiments, numerical simulations assume a key role in supporting them by finding the best initial condition to bring the experiment to the optimal results. 
The current state of the art of the numerical simulation in plasma acceleration are the \emph{Particle-in-cell} (PIC) codes~\cite{birdsall2018plasma}: they are employed to solve the Maxwell equations and the motion of "macroparticles" represented by a statistical ensemble of plasma particles; the positions and moments of the "macroparticles" are the average of the ensemble in the 6N-dimensional phase space. In the last years, new physical effects have been included in PIC codes such as collisions~\cite{perez2012improved,derouillat2018smilei}, ionization~\cite{nuter2011field,derouillat2018smilei}, warm plasma effects~\cite{birdsall2018plasma} and QED effects~\cite{arber2015contemporary}; however, the large number of particles of the plasma leads to high computational cost for this kinetic approach and the introduction of the macroparticles brings to numerical noise~\cite{birdsall2018plasma}.
An alternative to PIC codes are those based on \emph{fluid models}~\cite{birdsall2018plasma} that look at the plasma as a charged fluid and solve the equations for the macroscopic quantities (i.e., density and moments). In this way, there is a large spare of computational cost at the expense of physical effects reproduced by the codes~\cite{massimo2016comparisons}; fluid models for plasma acceleration are traditionally based on the \emph{Cold Fluid} (CF) approximation, used to close the hydrodynamic equations obtained from the Vlasov-Maxwell system. \\
In this article, we reproduce a PWFA process with a novel simulation route based on the \textit{Lattice Boltzmann Method} (LBM)~\cite{benzi1992lattice,succi2018lattice,kruger2017lattice}. The LBM is well suited to simulate complex problems of hydrodynamics starting from a discrete kinetic theory and modern developments find various successfully applications in other fields~\cite{succi2018lattice,kruger2017lattice}. We will work in the framework of the hybrid PIC/fluid code Architect~\cite{marocchino2016efficient} that employs a CF model solved with the {\it Flux Corrected Transport} (FCT) scheme~\cite{boris1973flux}. This CF implementation is replaced by the LBM implementation that allows to obtain a fluid model where diffusion effects coming from collisions between particles can be tuned by a specific parameter. The LBM implementation is constructed in such a way that in the limit of small diffusivity the CF model is recovered. We will then study the effects that a finite diffusivity produces on the profiles and the size of the electron bubble. Our study also shows that the LBM is a computational fluid solver that is more efficient than the FCT fluid solver implemented in Architect.\\
This paper is organized as follows: Sec.~\ref{sec:Arch} presents a short description of the hybrid kinetic-fluid approach with a focus on the hybrid PIC/fluid code Architect and the CF equations; Sec.~\ref{sec:LBM} introduces the LBM main equations and the resulting hydrodynamic limit that is compared with the CF equations; numerical results will be presented in Sec.~\ref{sec:Results}; conclusions will follow in Sec.~\ref{sec:conclusions}.
\section{Hybrid kinetic-fluid Approach}\label{sec:Arch}
Architect is a code based on a hybrid PIC/fluid approach used to simulate PWFA processes. It treats the evolution of electron bunches, driver or witness, as a 3D-PIC code~\cite{birdsall2018plasma}; the computational macroparticles move in a 6N-dimensional phase space, three for the positions and three for the moments. The electron plasma is treated as a cold, relativistic fluid~\cite{marocchino2016efficient} while plasma ions do not evolve. For the plasma fluid, cylindrical symmetry is assumed. In this scheme, the coupling between the plasma and the particle bunch is granted by the electromagnetic fields computed from the currents projected on a grid. This approach allows us to catch more physical phenomena in the bunch evolution but with a high spare of computational time because of the reduced model used for the plasma evolution~\cite{massimo2016comparisons}.

The equations describing the macroparticles motion of the bunch in the PIC subroutine of Architect are:

\begin{equation}
\begin{split}
&\frac{d \mathbf{x}_b}{d t} = c\boldsymbol{\beta}_b~,\\
&\frac{d\mathbf{p}_b}{d t}=-e(\mathbf{E}+c\boldsymbol{\beta}_b \times \mathbf{B})~,
\label{eq:bunch}
\end{split}
\end{equation}
where $d/dt$ is the total time derivative, $\mathbf{x}_b$ and $\mathbf{p}_b$ are respectively the position vector and the momentum vector of the macroparticles, $c$ is the speed of light, $\boldsymbol{\beta}_b = \mathbf{v}_b/c$ is the normalized velocity, $e$ is the electron charge, $\mathbf{E}$ and $\mathbf{B}$ are the electric and magnetic field vectors, respectively.

The equations for the plasma evolution are the so-called CF equations obtained from the Vlasov equation neglecting the pressure contribution (i.e., the thermal effects):
\begin{equation}\label{eq:plasma}
    \begin{split}
    &\frac{\partial n_e}{\partial t} + \boldsymbol{\nabla}_\mathbf{x} \cdot \left(n_e c\boldsymbol{\beta}_e \right) = 0~,\\
     &\frac{\partial \mathbf{p}_e}{\partial t} +c\boldsymbol{\beta}_e \cdot\boldsymbol{\nabla}_\mathbf{x} \mathbf{p}_e = -e\left(\mathbf{E} +c\boldsymbol{\beta}_e\times\mathbf{B}\right)~,\\
     & \boldsymbol{\beta}_e=\frac{\mathbf{p}_e}{m_e c\sqrt{1+|\mathbf{p}_e/m_e c|^2}}~,
\end{split}
\end{equation}
where $n_e$ is the electron number density and $m_e$ is the electron mass. We can rewrite these equations considering the cylindrical symmetry in the following form:
\begin{equation}\label{eq:continuitycoldfluid}
    \frac{\partial A}{\partial t} + \boldsymbol{\nabla}_\mathbf{x} \cdot (A\mathbf{u}) + \frac{u_r A}{r}= F~,
\end{equation}
where the quantities $A$, $\mathbf{u}$ and $F$ are defined according to the quantities of the Eq.~\eqref{eq:plasma}:
\begin{equation}\label{eq:plasma_quantities}
    A=\begin{bmatrix}n_e\\ n_e p_z\\ n_e p_r\end{bmatrix}~,~F=\begin{bmatrix} 0\\ -e n_e\left(\mathbf{E} +c\boldsymbol{\beta}_e\times\mathbf{B}\right)_z\\ -e n_e\left(\mathbf{E} +c\boldsymbol{\beta}_e\times\mathbf{B}\right)_r\end{bmatrix}~,~
    \mathbf{u}=c\boldsymbol{\beta}_e~.
\end{equation}
Eq.~\eqref{eq:continuitycoldfluid} represents three transport equations where $A$ is the scalar transported by the velocity field $\mathbf{u}$ and $F$ is the source term: it is first solved with the FCT~\cite{boris1973flux} scheme for the fluid advection part and then the macroparticles motion is computed through the electromagnetic fields (i.e., the Lorentz force) as a source term. This source term is computed by deposing the charge densities and the currents on the grid. The boundary conditions are symmetrical on the z-axis according to the cylindrical symmetry~\cite{birdsall2018plasma} while free flux boundary conditions~\cite{birdsall2018plasma} are implemented on the other edges.
The electromagnetic fields are integrated via Faraday law and Ampère-Maxwell equations:
\begin{equation}\label{eq:Maxwell}
    \begin{split}
    &\frac{d \mathbf{B}}{dt}=\boldsymbol{\nabla}\times\mathbf{E}~,\\
    &\frac{d \mathbf{E}}{d t}= c^2\boldsymbol{\nabla}\times\mathbf{B} +e\mu_0 c^3\left(\mathbf{J}_e+\mathbf{J}_b\right)~,\\
    &\mathbf{J}_e=n_e\boldsymbol{\beta}_e~,\\
    &\mathbf{J}_b=n_b\boldsymbol{\beta}_b~,
    \end{split}
\end{equation}
where $\mathbf{J}_e$ and $\mathbf{J}_b$ are the plasma electrons current and the bunch current respectively, $n_b$ is the bunch numerical density. In this way,  Eqs.~\eqref{eq:bunch} and~\eqref{eq:plasma} are coupled. Eqs.~\eqref{eq:Maxwell} are solved with the Finite Difference Time Domain (FDTD)~\cite{yee1966numerical} scheme. The boundary conditions are implemented as for the hydrodynamic fields according to the cylindrical symmetry on the z-axis~\cite{birdsall2018plasma} and free flux conditions are implemented on the other edges~\cite{birdsall2018plasma}.
The macroparticles of the bunch are computed on a 6N-dimensional phase space, the plasma and the electromagnetic fields are computed on a z-r grid (assuming a cylindrical symmetry) and then they are projected over the entire 6N-dimensional phase space to compute the force over the macroparticles. Moreover, to spare computational time, a moving window technique is implemented, so the plasma and electromagnetic quantities are integrated on a grid that moves together with the center of mass of the particles bunch.
\begin{figure*}[ht!]
    \centering
    \includegraphics[width=.98\linewidth]{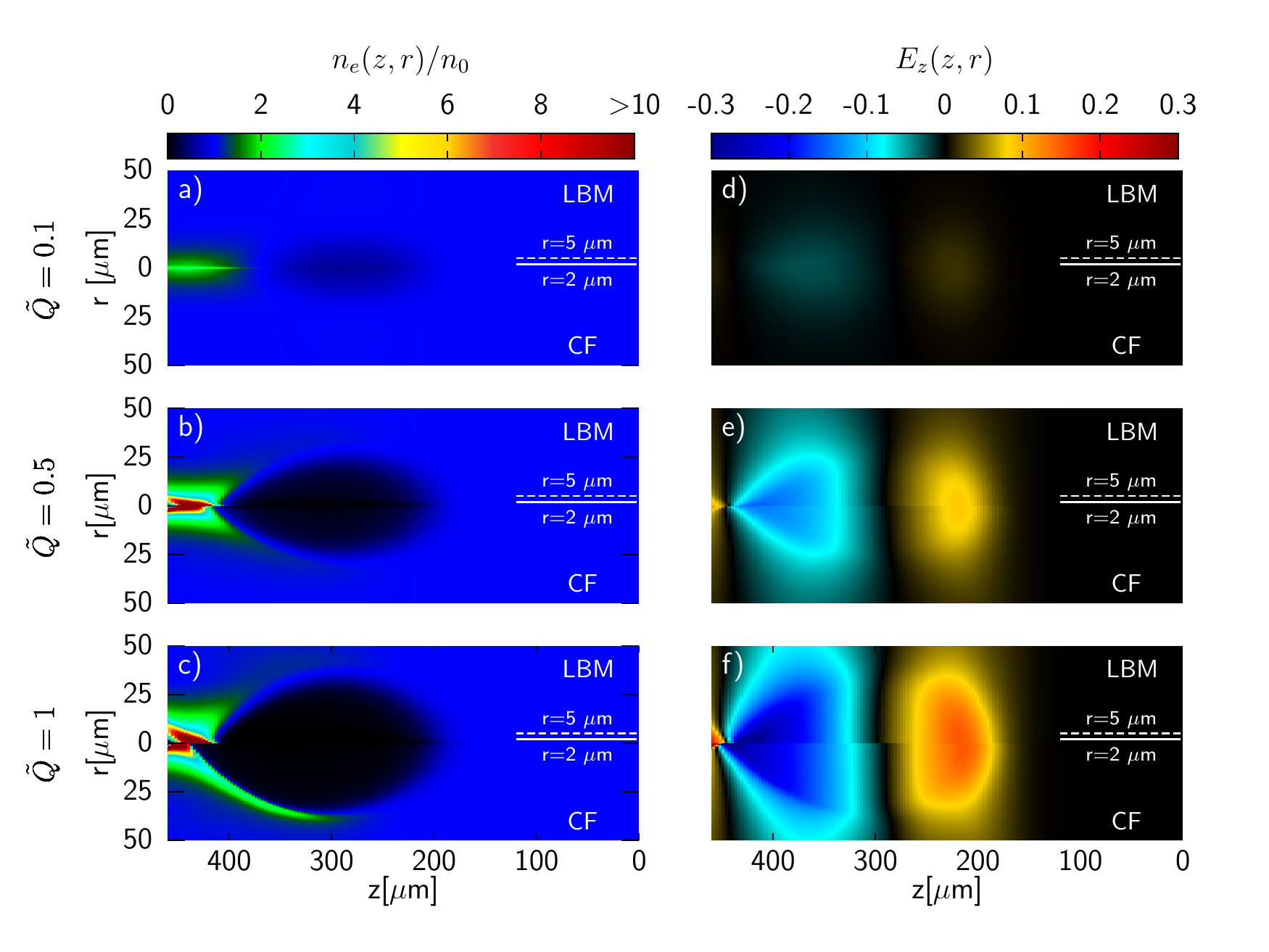}
    \caption{The 2D colour maps in the $z-r$ plane for the plasma electron density $n_e$ (panels a-c) and for the longitudinal component of the electric field $E_z$ (panels d-f) for three different values of $\tilde{Q}$ after $1~\mbox{mm}$ of propagation of the driver bunch into the plasma moving from left to right. In each panel, we report both data from LBM simulations (top-half side) and from CF simulations (bottom-half side). To quantitatively compare data from LBM and CF simulations at fixed values of $r$ near the center (i.e., $r=0$), two values of $r$ have been considered  (see Figs.~\ref{fig:ne_cut}~and~\ref{fig:Ez_cut}): $r=2~\mu\mbox{m}$ (continuous segment) and $r=5~\mu\mbox{m}$ (dashed segment). These comparisons are illustrated in Figs.~\ref{fig:ne_cut}-\ref{fig:Ez_cut}.}
    \label{fig:qualitative}
\end{figure*}
%

\begin{figure*}[ht!]
    \centering
    \includegraphics[width=.98\linewidth]{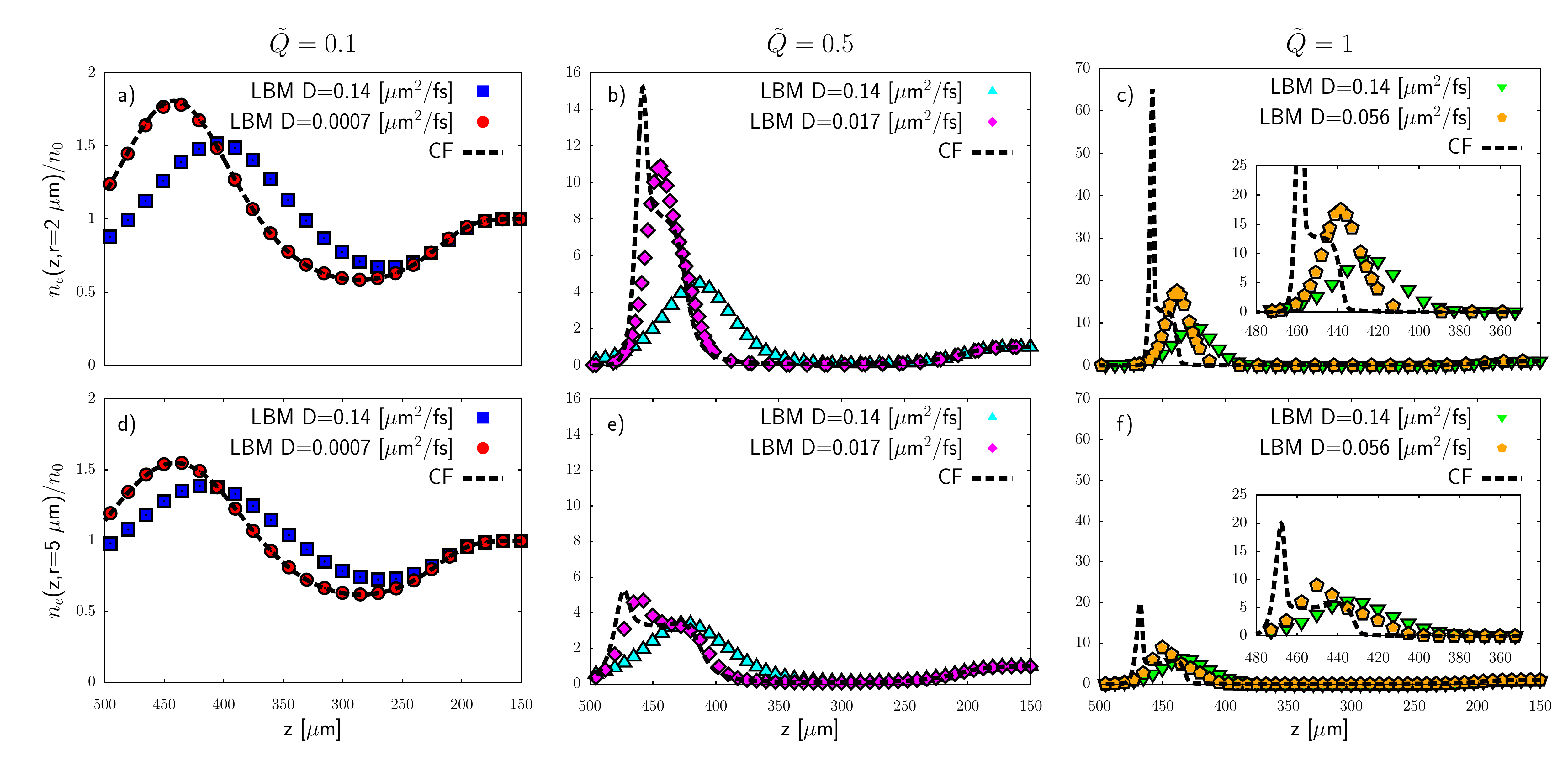}
    \caption{The profile of the plasma electron density for $r=2~\mu\mbox{m}$ (panels a-c) and $r=5~\mu\mbox{m}$ (panels d-f) are reported for three different values of $\tilde{Q}$ after $1~\mbox{mm}$ of propagation of the driver bunch into the plasma moving from left to right. In each panel, we compare the data from the LBM simulations for different values of the diffusivity (filled points) with data from the CF simulations (dashed black line).}
    \label{fig:ne_cut}
\end{figure*}
\begin{figure*}[ht!]
    \centering
    \includegraphics[width=.98\linewidth]{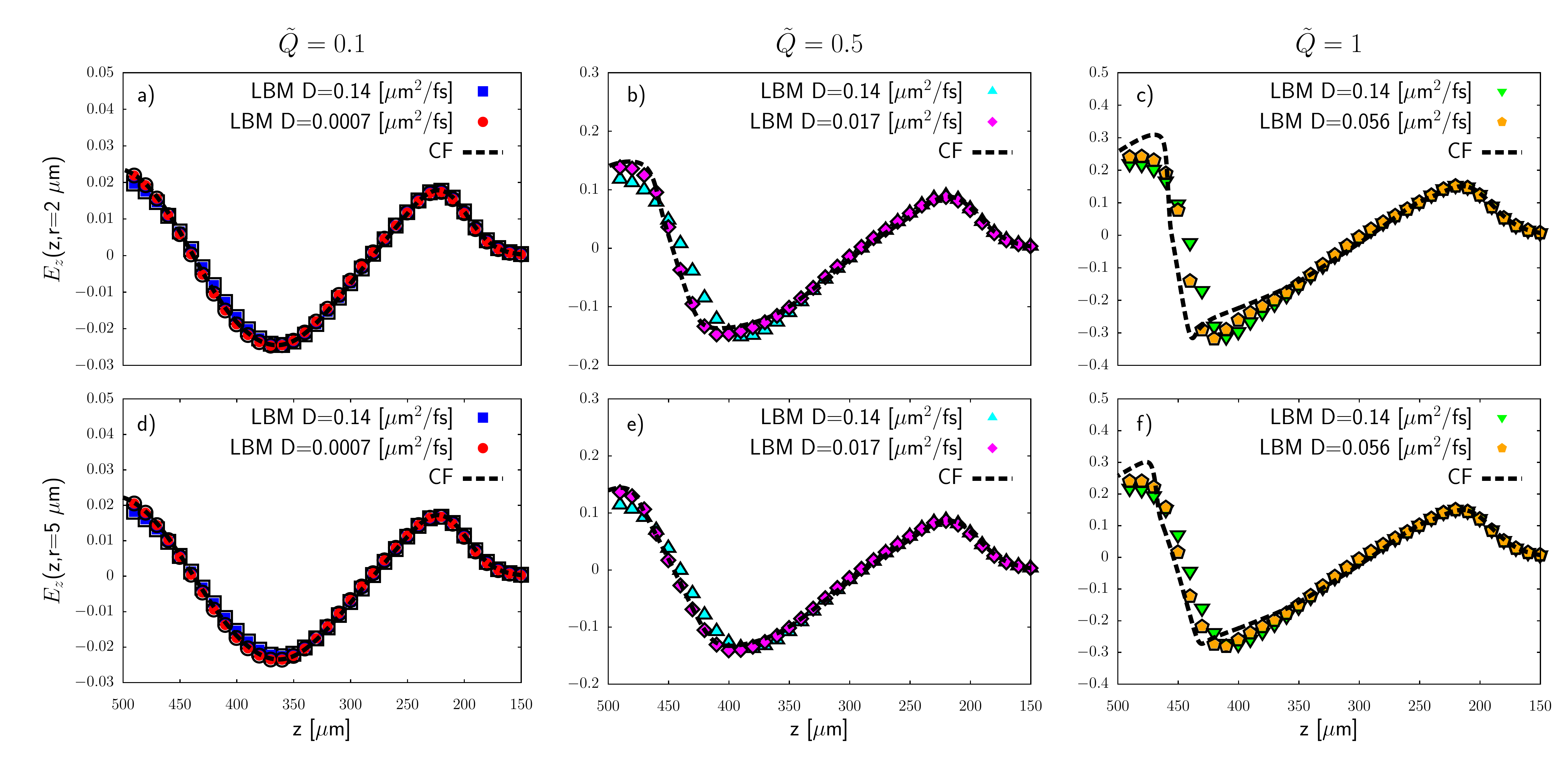}
    \caption{The profile of the longitudinal electric field for $r=2~\mu\mbox{m}$ (panels a-c) and $r=5~\mu\mbox{m}$ (panels d-f) are reported for three different values of $\tilde{Q}$ after $1~\mbox{mm}$ of propagation of the driver bunch into the plasma from left to the right. In each panel we compare the LBM simulations with different values of the diffusivity (filled points) with the CF simulations (dashed black line). }
    \label{fig:Ez_cut}
\end{figure*}
%
\section{Lattice Boltzmann Method}\label{sec:LBM}
\begin{figure}[ht!]
    \centering
        \includegraphics[width=.7\linewidth,]{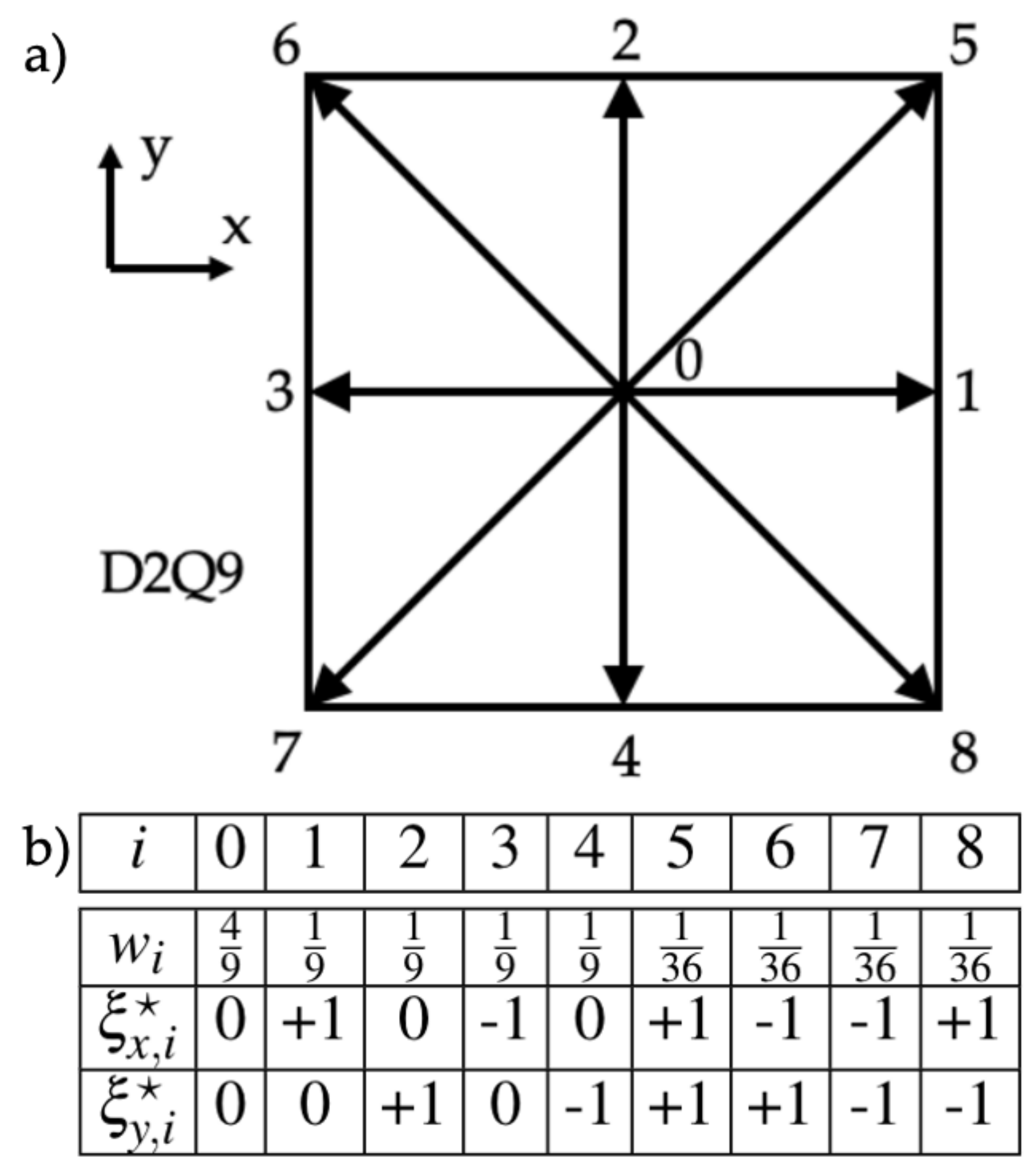}
    \caption{Panel a): a D2Q9-LBM velocity scheme. Panel b): the explicit set of adimensional discrete velocities $\boldsymbol{\xi}_i^\star=\boldsymbol{\xi}_i/(\Delta x/\Delta t)$ with the corresponding statistical weights $w_i$.}\label{fig:D2Q9}
\end{figure}

In this section, we recall the basic features of the LBM that we adopted in our study. Our strategy is to recover the hydrodynamic description in Eqs.~\eqref{eq:plasma} supplemented by diffusion effects. The PIC subroutine in Architect employed to solve Eqs.~\eqref{eq:bunch} and the integration of the electromagnetic fields in Eqs.~\eqref{eq:Maxwell} will not be changed.\\
The LBM is grounded in kinetic theory and studies the evolution of a kinetic probability distribution function $f_i(\mathbf{x},t)$ to find a ``fluid particle" in the position $\mathbf{x}$, at the time $t$ with a kinetic velocity $\boldsymbol{\xi}_i$. Space is discretized on a regular grid while time advances with a finite interval $\Delta t$. Moreover, the kinetic velocities are discretized as well ($i=0,1,...,N_{\mbox{\tiny tot}}-1$) and are defined by a finite set designed to properly recover the desired hydrodynamic description~\cite{benzi1992lattice,chen1998lattice,succi2018lattice,kruger2017lattice}. One of the key advantages of the method is the ability to provide a hydrodynamic description with a very limited set of kinetic velocities. In this work, we use the so-called D2Q9~\cite{kruger2017lattice} scheme for a 2D grid with nine kinetic velocities ($N_{\mbox{\tiny tot}}=9$) (see Fig/Table~\eqref{fig:D2Q9}). 
The evolution of the probability distribution function is computed according to the discrete Boltzmann equation~\cite{benzi1992lattice,kruger2017lattice}:
\begin{equation}
f_i(\mathbf{x}+\boldsymbol{\xi}_i\Delta t,t+\Delta t) - f_i(\mathbf{x},t)= \Omega_c(\mathbf{x},t) \Delta t + S_i(\mathbf{x},t)~,
\label{eq:boltzmann}
\end{equation}
where $\Omega_c(f_i)$ is the collision term and $S_i$ accounts for the source term; in this work, we chose the BGK collision term~\cite{bhatnagar1954model}, expressing the relaxation of the probability distribution towards a local equilibrium: 
\begin{equation}\label{eq:BGK}
\Omega_c(\mathbf{x},t) = -\frac{f_i(\mathbf{x},t)-f_i^{\mbox{\tiny (eq)}}(\mathbf{x},t)}{\tau}~,\\
\end{equation}
where $f_i^{\mbox{\tiny (eq)}}(\mathbf{x},t)$ is the equilibrium distribution function that can be chosen to reproduce the desired hydrodynamics, while $\tau$ is the relaxation time, that is, the time that the present probability distribution takes to relax to the local equilibrium. In the framework of the LBM, the hydrodynamic fields are given by the zeroth- and the first-order moments of the probability distribution functions $f_i(\mathbf{x},t)$:
\begin{equation}
    \begin{split}
        A(\mathbf{x},t)&=\sum_{i=0}^{N_{\mbox{\tiny tot}}-1} f_i(\mathbf{x},t)~,\\
        A(\mathbf{x},t)\mathbf{u}(\mathbf{x},t)&= \sum_{i=0}^{N_{\mbox{\tiny tot}}-1} \boldsymbol{\xi}_i f_i(\mathbf{x},t)~,\\        \label{eq:hd_fields}
    \end{split}
\end{equation}
where $A$ is a scalar and $\mathbf{u}$ is the advecting velocity field. The hydrodynamic description is mathematically granted by the Chapman-Enskog expansion analysis~\cite{benzi1992lattice,kruger2017lattice} and depends on the choice of the local equilibrium distribution function $f_i^{\mbox{\tiny (eq)}}$ in Eq.~\eqref{eq:BGK}. We choose the following equilibrium distribution function which depends on $\mathbf{x}$ and $t$ via the scalar field $A=A(\mathbf{x},t)$ and the velocity field $\mathbf{u}=\mathbf{u}(\mathbf{x},t)$:
\begin{equation}
        f_i^{\mbox{\tiny (eq)}}(A,\mathbf{u})=w_{i} A \left(1+\frac{\mathbf{u} \cdot \boldsymbol{\xi}_{i}}{3}+\frac{(\mathbf{u}\cdot \boldsymbol{\xi}_{i})^{2}}{18}-\frac{\mathbf{u}\cdot \mathbf{u}}{6} \right)~,
\end{equation}
where $w_i$ are statistical weights whose values are reported in Fig.~\ref{fig:D2Q9}b. Concerning the source term, we adopt the following choice~\cite{kruger2017lattice}:
\begin{equation}
S_{i}(\mathbf{x},t)= w_{i}F(\mathbf{x},t)~,
\end{equation}
where $F(\mathbf{x},t)=\sum_{i=0}^{N_{\mbox{\tiny tot}}-1} S_i(\mathbf{x},t)$ is the source term that will result in the dynamical equation for the scalar $A(\mathbf{x},t)$.\\
The hydrodynamic description emerging from the above described LBM corresponds to an advection-diffusion equation~\cite{kruger2017lattice,succi2018lattice}:
\begin{equation}\label{eq:A-D}
\frac{\partial A}{\partial t} + \boldsymbol{\nabla}_\mathbf{x}\cdot(A\mathbf{u})=D\boldsymbol{\nabla}_\mathbf{x}^2 A + F\ ,
\end{equation}
where the diffusion parameter $D$ is given by 
\begin{equation}\label{eq:diff_par}
D=\frac{1}{3}\left(\tau-\frac{\Delta t}{2}\right)\ ,
\end{equation}
and can be tuned via the choice of $\tau$ in the lattice Boltzmann equation (see Eq.~\eqref{eq:BGK}). Since the code Architect implements the approximation of cylindrical symmetry, we adapted the LBM to this framework following refs.~\cite{srivastava2013axisymmetric,premnath2005lattice,zhou2011axisymmetric}. Using cylindrical coordinates means adding an additional source term to the discrete Boltzmann equation  (see Eq.~\eqref{eq:boltzmann}) that represents the additional factors coming from the change of coordinates. Moving to cylindrical coordinates $(z,r,t)$, two terms are added to Eq.~\eqref{eq:A-D} coming from the gradient and Laplace operators:
\begin{equation}
    \frac{\partial A}{\partial t} + \boldsymbol{\nabla}_\mathbf{x} \cdot (A\mathbf{u}) = D \left(\boldsymbol{\nabla}_\mathbf{x}^2 A + \frac{1}{r} \nabla_r{A}\right)+ F - \frac{A u_r}{r}.
    \label{eq:cylindrical A-D}
\end{equation}
To compare with the CF model (see Eq.~\eqref{eq:plasma}) we then use three LBM descriptions where the scalar field $A$, the advecting velocity $\mathbf{u}$ and the source terms are chosen in compliance with Eqs.~\eqref{eq:plasma_quantities}. On the top of this CF description, we can introduce diffusion effects via the diffusion coefficient $D$ that we can tune with the relaxation time $\tau$ as in~\eqref{eq:diff_par}.
The boundary conditions are constructed in such a way to reproduce the boundary conditions introduced in Sec.~\ref{sec:Arch}: the probability distribution function on the boundary $f_{i}$ is equal to an equilibrium distribution function:
\begin{equation}
    f_{i}(\mathbf{\ubar{x}},t)=f_{i}^{\mbox{\tiny (eq)}}(A(\mathbf{\ubar{x}},t),\mathbf{u}(\mathbf{\ubar{x}},t))\ ,
\end{equation}
where $\mathbf{\ubar{x}}$ is the boundary position and $A(\mathbf{\ubar{x}},t)$ is the scalar field on the boundary computed according to the cylindrical symmetry on the z-axis~\cite{birdsall2018plasma} and to the free flux boundary on the other edges~\cite{birdsall2018plasma}.
\section{Results}\label{sec:Results}
In this section, we present the results obtained with the LBM implementation for the plasma evolution and compare them with the results of a CF model solved with the FCT scheme~\cite{boris1973flux}. We perform comparisons in the linear, quasi-non-linear and non-linear regimes by choosing different normalized charge~\cite{rosenzweig1987nonlinear} parameters defined as:
\begin{equation}
    \Tilde{Q}=(2\pi)^{3/2}\sigma_x \sigma_y \sigma_z \alpha k_p^{-3}\ ,
    \label{eq:Qtilde}
\end{equation}
where $\sigma_i~(i=x,y,z)$ are the rms-sizes of the driver in the corresponding direction, $\alpha=\rho_b/\rho_0$ is the ratio between the charge density of the bunch and the unperturbed plasma charge density and $k_p=\omega_p/c$ is the plasma wave number ($\omega_p=\sqrt{4\pi e^2 n_e/m_e}$). To test the codes, we simulate the propagation of a bi-Gaussian electron bunch in an unperturbed plasma with numerical density $10^{16}~\mbox{cm}^{-3}$. The bunch parameters are: $2\times 10^5$ particles, $\sigma_z=17.7~\mu\mbox{m}$, $\sigma_x=\sigma_y=8.9~\mu\mbox{m}$, Lorentz factor $\gamma=5\times10^3$, transverse emittance $\varepsilon_x=\varepsilon_y=74~\mbox{mm~mrad}$, relative energy spread $\delta \gamma/\gamma=0.01\%$ and a charge of $(0.024,0.120,0.240)~\mbox{nC}$ corresponding to a normalized charge parameter of $\Tilde{Q}=0.1,0.5,1$. The driver center of mass is fixed in the position $z=200~\mu\mbox{m},~r=0~\mu\mbox{m}$ and is not showed because we are interested in the plasma evolution. The grid parameters are: length $600~\mu\mbox{m}$, transverse length $100~\mu\mbox{m}$, cell dimension $\Delta z= \Delta r=0.25~\mu\mbox{m}$  and a computational time step of $\Delta t_c =0.147~\mbox{fs}$. A first qualitative comparison is given in Fig.~\ref{fig:qualitative}, where we compare the  2D colour map of the density $n_e$ and the longitudinal electric field $E_z$; the top-half of the plot reports the LBM result while the bottom-half reports the CF result. For the LBM we fix the value of the diffusion coefficient $D$ to be the smallest value that guarantees numerical stability (see also discussion below). The two methods qualitatively produce similar results: increasing the value of $\Tilde{Q}$ results in an increase of the size of the so-called electron bubble, of the density peak at the back of the bubble and of the intensity of the electromagnetic field. 
The fluid dynamic description of LBM is intrinsically affected by the diffusivity induced by the relaxation time (see Eq.~\eqref{eq:diff_par}), which prompts a quantitative comparison between the LBM results and CF results, changing the diffusivity $D$. To this aim, we take plots of both the density $n_e$ and the longitudinal electric field $E_z$ along the longitudinal coordinate $z$ for fixed values of the radial coordinate, $r= 2\,\mu m$ and $r= 5\,\mu m$. The first value is chosen to compare the methods near the axis without boundary effects and the second value is chosen to compare the results away from the axis. We perform LBM simulations for two representative values of the diffusivity to highlight its effect on the profiles. Results are reported in Figs.~\ref{fig:ne_cut} and~\ref{fig:Ez_cut}. In the linear regime, Fig.~\ref{fig:ne_cut} panels a) and d), we can correctly reproduce the CF results in the limit of a very small diffusivity. Approaching the non-linear regime, however, the smallest $D$ value allowing for numerical stability of the LBM increases. This problem could arise because, in the non-linear regime, gradients of the fields are more pronounced and the collisional operator $\Omega_c$ in Eq.~\eqref{eq:BGK} does not grant a sufficient numerical stability for such situations. We note that our choice of $\Omega_c$ is the simplest one in the LBM framework and there are other (more complicated) choices~\cite{kruger2017lattice,succi2018lattice} that could be tested to improve the capability of the LBM simulations in these regimes. This surely warrants future investigations. Results in Fig.~\ref{fig:ne_cut}, panels b) and e), clearly show how the effects of diffusion act on the electron density profile, resulting in a reduction of the size of the electron bubble; these aspects are even more amplified in Fig.~\ref{fig:ne_cut}, panels c) and f): the bubble simulated with the LBM is about 10\% shorter then the bubble simulated with the CF Eq.~\eqref{eq:plasma}. The CF equations predict a singularity ($n_e\rightarrow\infty$) behind the tail of the bubble in the limit of \emph{cold wave breaking} and diffusivity actually acts as a regularization mechanism for these incipient singularities. Other studies in the literature already considered regularization effects in the dynamics of electron plasma waves using warm plasma equations with a non zero temperature in the limit of a 1D model~\cite{schroeder2005warm}, also including transverse fluctuations~\cite{schroeder2010relativistic}; in our case we are considering a 2D-axisymmetric case where regularization comes from diffusion in density-momentum space.  While a change in the diffusivity greatly affects the electron density profiles in the wake of the bubbles, the effects on the longitudinal electric fields are more mitigated in all regimes. We indeed observe in Fig.~\ref{fig:Ez_cut} that the longitudinal electric field keeps a good agreement with the results of the CF simulations for all the simulated values of $\tilde{Q}$. This can be attributed to the fact that the peak of density predicted by the CF equation is a local effect owned by Eq.~\eqref{eq:plasma} that does not affect globally the Eq.~\eqref{eq:Maxwell} of the electromagnetic fields.
\begin{figure}[!ht]
    \centering
    \includegraphics[width=1.\linewidth]{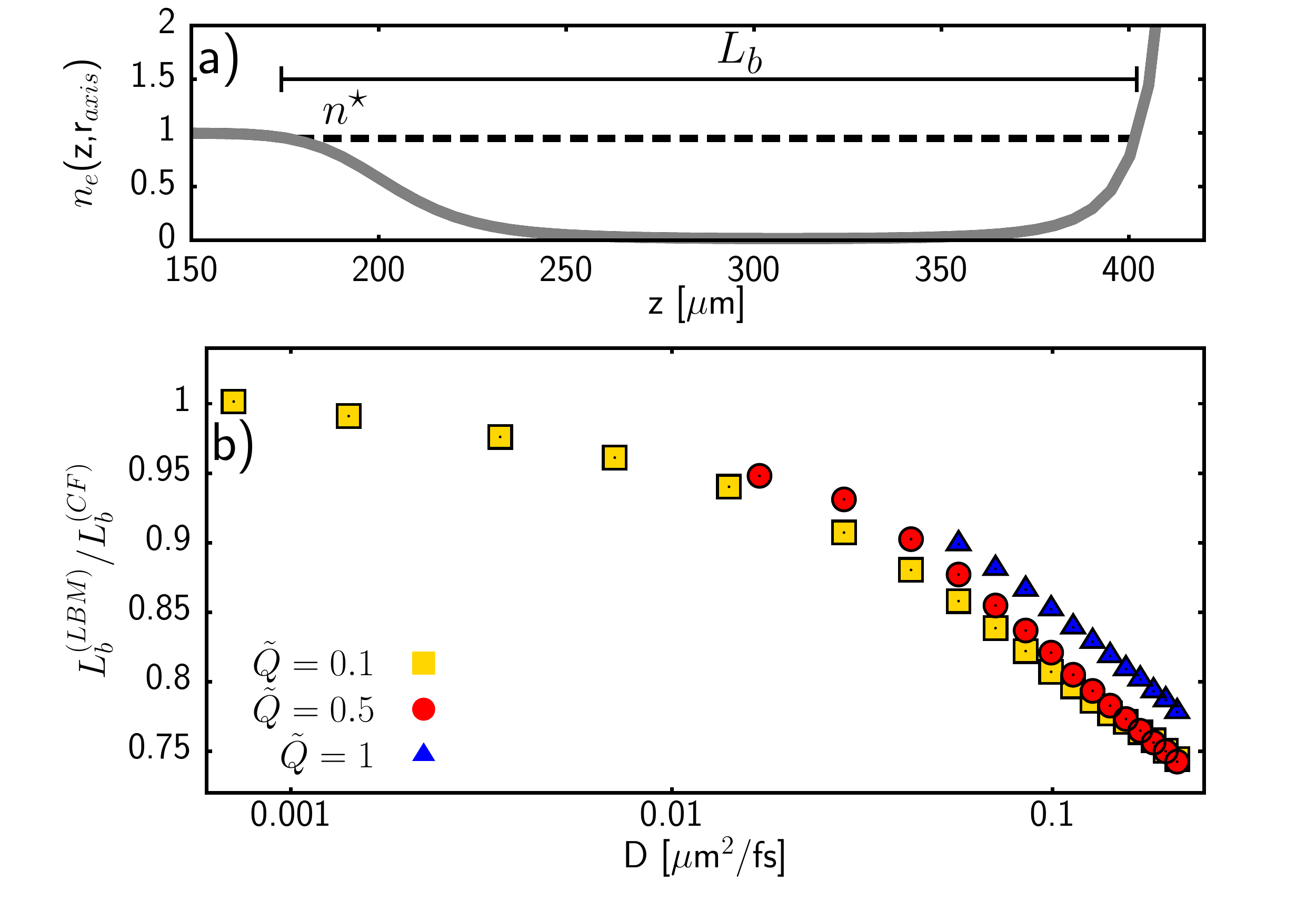}
    \caption{Panel a): sketch of the method used to compute the bubble length $L_b$ in terms of the threshold value $n^\star=0.95$ on the electron density profile. Panel b): we report the ratio between the bubble length obtained in the LBM simulations and in the CF simulations as a function of the diffusivity.}
    \label{fig:bubble_lenght}
\end{figure}

In Fig.~\ref{fig:bubble_lenght}, we look at the bubble length $L_b$. We retrieved it by choosing a threshold value of the electron density $n^\star=0.95$ (see Fig.~\ref{fig:bubble_lenght}, panel a)) and counted the number of sites with a lower density; by multiplying the number of sites for the size cell $\Delta z$, we find the value of $L_b$ in micron. The bubble length $L_b^{(LBM)}$ obtained via the LBM  increases as the diffusion parameter $D$ decreases and tends asymptotically to the bubble length $L_b^{(CF)}$ obtained from simulations of the CF (see Eq.~\eqref{eq:plasma}). This behaviour shows the tendency of the electrons to restore density equilibrium after the perturbation of the driver in a faster way if the diffusion coefficient $D$ is larger; this results in a shorter bubble.\\ 
We finally analyzed the performance of the fluid solver based on the LBM in comparison to the fluid solver based on the FCT scheme~\cite{boris1973flux}. In Tab.~\ref{tab:efficiency}, we report the computational time spent by the FCT and LBM routines. This comparison is performed for three different resolutions and highlights a spare of computational time when using the LBM routine. The FCT method is a finite differences method that makes a weighted average between a lower order scheme and an higher order scheme~\cite{zalesak1979fully} to integrate the Eqs.~\eqref{eq:plasma} increasing the precision of the results at the expenses of the computational cost; on the other hand, LBM hinges on Eq.~\eqref{eq:boltzmann}, which is a hyperbolic equation for the evolution of the probability distribution function $f$ to recover the hydrodynamics fields.
\begin{table}[h!]
    \centering
    \begin{tabular}{m{2.1cm} m{2.1cm} m{2.1cm} m{1.8cm}}
     \hline
     Number of cells & FCT [s$\times10^3$] & LBM [s$\times10^3$] & Gain \%\\
     \hline
     \hline
     $6\times10^4$ & $0.57$ & $0.34$ & $40.35\%$\\
     $24\times10^4$ & $9.08$ & $3.03$ & $66.63\%$\\
     $96\times10^4$ & $107.55$ & $28.88$ & $73.15\%$\\
     \hline
\end{tabular}
    \caption{Comparison of computational efficiency of FCT and LBM routine to simulate 2 mm of propagation of the driver in the plasma at changing the size of the computational grid. The Gain column is computed with ((FCT-LBM)/FCT)\%. Simulations are made on Intel Xeon E5-2695@2.40GHz.}
    \label{tab:efficiency}
\end{table}
%
\section{Conclusions}\label{sec:conclusions}
In this article, we reported results of Plasma Wakefield Acceleration (PWFA) simulations performed with the Lattice Boltzmann Method (LBM), a hydrodynamic solver well known in the literature~\cite{benzi1992lattice,chen1998lattice,succi2018lattice,kruger2017lattice} that has never been used before to simulate a plasma acceleration process. We started from the code Architect~\cite{marocchino2016efficient}, implementing a Cold Fluid (CF) model for the plasma and then adapted the LBM in order to simulate the fluid part of the code. Importantly, the LBM introduces diffusion effects in the plasma evolution, thus going beyond the CF approximation. The results of the simulations support the applicability of the LBM up to the onset of the non-linear regime. The future developments of the present study are multifold. We found that LBM is more efficient than the FCT method that is used to reproduce the CF model in Architect, hence a 3D numerical simulation and/or longer 2D simulations are doable with a reasonable computational time. This could be useful to address important physical effects, such as the hose instability in 3D~\cite{mehrling2018accurate} or the physics emerging in the study of the linear plasma colliders~\cite{adli2019plasma}, where many bunches are injected into the plasma with a high repetition rate. Regarding the role of the diffusivity that is peculiar of LBM, we note that the diffusion effects have not been considered until now because they are small in the early periods of the plasma wave, but they may become important, for example, in the high repetition rate studies. These observations call for a precise determination of the diffusion term, that will be the main target for forthcoming theoretical and experimental studies. \\
More generally, our computational model inscribes in the framework of those approaches that go beyond the CF approximation but still retain some hydrodynamic characters. In this context, an earlier attempt has already been made to quantify thermal  effects in a relativistic warm fluid models~\cite{schroeder2010relativistic,schroeder2005warm}. Along these lines, it could be of interest to perform numerical simulations with relativistic LBM~\cite{gabbana2020relativistic} that is able to reproduce the hydrodynamics of 3D relativistic fluids. Considering the numbers of projects, like FACET~\cite{yakimenko2019facet}, EuPRAXIA~\cite{assmann2020eupraxia}, FlashForward~\cite{aschikhin2016flashforward} among the others, devoted to produce high quality electron beam by using PWFA, a simulation tool able to calculate properly and faster the features of plasma acceleration is very attractive. Moreover, even people working on high energy physics, committed to develop future compact linear colliders, can benefit from a code that goes beyond the CF approximation, paving the way for the simulations of the high repetition rate machines, which are necessary for this purpose.
\begin{acknowledgments}
We acknowledge fruitful discussions with Jamie Rosenzweig, Carlo Benedetti, Fulvio Zonca, Paolo Santangelo.
\end{acknowledgments}

\nocite{*}
\bibliography{ref.bib}

\begin{thebibliography}{32}%
\makeatletter
\providecommand \@ifxundefined [1]{%
 \@ifx{#1\undefined}
}%
\providecommand \@ifnum [1]{%
 \ifnum #1\expandafter \@firstoftwo
 \else \expandafter \@secondoftwo
 \fi
}%
\providecommand \@ifx [1]{%
 \ifx #1\expandafter \@firstoftwo
 \else \expandafter \@secondoftwo
 \fi
}%
\providecommand \natexlab [1]{#1}%
\providecommand \enquote  [1]{``#1''}%
\providecommand \bibnamefont  [1]{#1}%
\providecommand \bibfnamefont [1]{#1}%
\providecommand \citenamefont [1]{#1}%
\providecommand \href@noop [0]{\@secondoftwo}%
\providecommand \href [0]{\begingroup \@sanitize@url \@href}%
\providecommand \@href[1]{\@@startlink{#1}\@@href}%
\providecommand \@@href[1]{\endgroup#1\@@endlink}%
\providecommand \@sanitize@url [0]{\catcode `\\12\catcode `\$12\catcode
  `\&12\catcode `\#12\catcode `\^12\catcode `\_12\catcode `\%12\relax}%
\providecommand \@@startlink[1]{}%
\providecommand \@@endlink[0]{}%
\providecommand \url  [0]{\begingroup\@sanitize@url \@url }%
\providecommand \@url [1]{\endgroup\@href {#1}{\urlprefix }}%
\providecommand \urlprefix  [0]{URL }%
\providecommand \Eprint [0]{\href }%
\providecommand \doibase [0]{http://dx.doi.org/}%
\providecommand \selectlanguage [0]{\@gobble}%
\providecommand \bibinfo  [0]{\@secondoftwo}%
\providecommand \bibfield  [0]{\@secondoftwo}%
\providecommand \translation [1]{[#1]}%
\providecommand \BibitemOpen [0]{}%
\providecommand \bibitemStop [0]{}%
\providecommand \bibitemNoStop [0]{.\EOS\space}%
\providecommand \EOS [0]{\spacefactor3000\relax}%
\providecommand \BibitemShut  [1]{\csname bibitem#1\endcsname}%
\let\auto@bib@innerbib\@empty
\bibitem [{\citenamefont {Litos}\ \emph {et~al.}(2014)\citenamefont {Litos},
  \citenamefont {Adli}, \citenamefont {An}, \citenamefont {Clarke},
  \citenamefont {Clayton}, \citenamefont {Corde}, \citenamefont {Delahaye},
  \citenamefont {England}, \citenamefont {Fisher}, \citenamefont {Frederico}
  \emph {et~al.}}]{litos2014high}%
  \BibitemOpen
  \bibfield  {author} {\bibinfo {author} {\bibfnamefont {M.}~\bibnamefont
  {Litos}}, \bibinfo {author} {\bibfnamefont {E.}~\bibnamefont {Adli}},
  \bibinfo {author} {\bibfnamefont {W.}~\bibnamefont {An}}, \bibinfo {author}
  {\bibfnamefont {C.}~\bibnamefont {Clarke}}, \bibinfo {author} {\bibfnamefont
  {C.}~\bibnamefont {Clayton}}, \bibinfo {author} {\bibfnamefont
  {S.}~\bibnamefont {Corde}}, \bibinfo {author} {\bibfnamefont
  {J.}~\bibnamefont {Delahaye}}, \bibinfo {author} {\bibfnamefont
  {R.}~\bibnamefont {England}}, \bibinfo {author} {\bibfnamefont
  {A.}~\bibnamefont {Fisher}}, \bibinfo {author} {\bibfnamefont
  {J.}~\bibnamefont {Frederico}},  \emph {et~al.},\ }\href@noop {} {\bibfield
  {journal} {\bibinfo  {journal} {Nature}\ }\textbf {\bibinfo {volume} {515}},\
  \bibinfo {pages} {92} (\bibinfo {year} {2014})}\BibitemShut {NoStop}%
\bibitem [{\citenamefont {Esarey}\ \emph {et~al.}(2009)\citenamefont {Esarey},
  \citenamefont {Schroeder},\ and\ \citenamefont
  {Leemans}}]{esarey2009physics}%
  \BibitemOpen
  \bibfield  {author} {\bibinfo {author} {\bibfnamefont {E.}~\bibnamefont
  {Esarey}}, \bibinfo {author} {\bibfnamefont {C.}~\bibnamefont {Schroeder}}, \
  and\ \bibinfo {author} {\bibfnamefont {W.}~\bibnamefont {Leemans}},\
  }\href@noop {} {\bibfield  {journal} {\bibinfo  {journal} {Reviews of modern
  physics}\ }\textbf {\bibinfo {volume} {81}},\ \bibinfo {pages} {1229}
  (\bibinfo {year} {2009})}\BibitemShut {NoStop}%
\bibitem [{\citenamefont {Blumenfeld}\ \emph {et~al.}(2007)\citenamefont
  {Blumenfeld}, \citenamefont {Clayton}, \citenamefont {Decker}, \citenamefont
  {Hogan}, \citenamefont {Huang}, \citenamefont {Ischebeck}, \citenamefont
  {Iverson}, \citenamefont {Joshi}, \citenamefont {Katsouleas}, \citenamefont
  {Kirby} \emph {et~al.}}]{blumenfeld2007energy}%
  \BibitemOpen
  \bibfield  {author} {\bibinfo {author} {\bibfnamefont {I.}~\bibnamefont
  {Blumenfeld}}, \bibinfo {author} {\bibfnamefont {C.~E.}\ \bibnamefont
  {Clayton}}, \bibinfo {author} {\bibfnamefont {F.-J.}\ \bibnamefont {Decker}},
  \bibinfo {author} {\bibfnamefont {M.~J.}\ \bibnamefont {Hogan}}, \bibinfo
  {author} {\bibfnamefont {C.}~\bibnamefont {Huang}}, \bibinfo {author}
  {\bibfnamefont {R.}~\bibnamefont {Ischebeck}}, \bibinfo {author}
  {\bibfnamefont {R.}~\bibnamefont {Iverson}}, \bibinfo {author} {\bibfnamefont
  {C.}~\bibnamefont {Joshi}}, \bibinfo {author} {\bibfnamefont
  {T.}~\bibnamefont {Katsouleas}}, \bibinfo {author} {\bibfnamefont
  {N.}~\bibnamefont {Kirby}},  \emph {et~al.},\ }\href@noop {} {\bibfield
  {journal} {\bibinfo  {journal} {Nature}\ }\textbf {\bibinfo {volume} {445}},\
  \bibinfo {pages} {741} (\bibinfo {year} {2007})}\BibitemShut {NoStop}%
\bibitem [{\citenamefont {Doyle}\ \emph {et~al.}(2019)\citenamefont {Doyle},
  \citenamefont {McDaniel},\ and\ \citenamefont {Hamm}}]{doyle2019future}%
  \BibitemOpen
  \bibfield  {author} {\bibinfo {author} {\bibfnamefont {B.~L.}\ \bibnamefont
  {Doyle}}, \bibinfo {author} {\bibfnamefont {F.~D.}\ \bibnamefont {McDaniel}},
  \ and\ \bibinfo {author} {\bibfnamefont {R.~W.}\ \bibnamefont {Hamm}},\
  }\href@noop {} {\bibfield  {journal} {\bibinfo  {journal} {Reviews of
  Accelerator Science and Technology}\ }\textbf {\bibinfo {volume} {10}},\
  \bibinfo {pages} {93} (\bibinfo {year} {2019})}\BibitemShut {NoStop}%
\bibitem [{\citenamefont {Chen}\ \emph {et~al.}(1985)\citenamefont {Chen},
  \citenamefont {Dawson}, \citenamefont {Huff},\ and\ \citenamefont
  {Katsouleas}}]{chen1985acceleration}%
  \BibitemOpen
  \bibfield  {author} {\bibinfo {author} {\bibfnamefont {P.}~\bibnamefont
  {Chen}}, \bibinfo {author} {\bibfnamefont {J.}~\bibnamefont {Dawson}},
  \bibinfo {author} {\bibfnamefont {R.~W.}\ \bibnamefont {Huff}}, \ and\
  \bibinfo {author} {\bibfnamefont {T.}~\bibnamefont {Katsouleas}},\
  }\href@noop {} {\bibfield  {journal} {\bibinfo  {journal} {Physical review
  letters}\ }\textbf {\bibinfo {volume} {54}},\ \bibinfo {pages} {693}
  (\bibinfo {year} {1985})}\BibitemShut {NoStop}%
\bibitem [{\citenamefont {Birdsall}\ and\ \citenamefont
  {Langdon}(2018)}]{birdsall2018plasma}%
  \BibitemOpen
  \bibfield  {author} {\bibinfo {author} {\bibfnamefont {C.~K.}\ \bibnamefont
  {Birdsall}}\ and\ \bibinfo {author} {\bibfnamefont {A.~B.}\ \bibnamefont
  {Langdon}},\ }\href@noop {} {\emph {\bibinfo {title} {Plasma physics via
  computer simulation}}}\ (\bibinfo  {publisher} {CRC press},\ \bibinfo {year}
  {2018})\BibitemShut {NoStop}%
\bibitem [{\citenamefont {P{\'e}rez}\ \emph {et~al.}(2012)\citenamefont
  {P{\'e}rez}, \citenamefont {Gremillet}, \citenamefont {Decoster},
  \citenamefont {Drouin},\ and\ \citenamefont {Lefebvre}}]{perez2012improved}%
  \BibitemOpen
  \bibfield  {author} {\bibinfo {author} {\bibfnamefont {F.}~\bibnamefont
  {P{\'e}rez}}, \bibinfo {author} {\bibfnamefont {L.}~\bibnamefont
  {Gremillet}}, \bibinfo {author} {\bibfnamefont {A.}~\bibnamefont {Decoster}},
  \bibinfo {author} {\bibfnamefont {M.}~\bibnamefont {Drouin}}, \ and\ \bibinfo
  {author} {\bibfnamefont {E.}~\bibnamefont {Lefebvre}},\ }\href@noop {}
  {\bibfield  {journal} {\bibinfo  {journal} {Physics of Plasmas}\ }\textbf
  {\bibinfo {volume} {19}},\ \bibinfo {pages} {083104} (\bibinfo {year}
  {2012})}\BibitemShut {NoStop}%
\bibitem [{\citenamefont {Derouillat}\ \emph {et~al.}(2018)\citenamefont
  {Derouillat}, \citenamefont {Beck}, \citenamefont {P{\'e}rez}, \citenamefont
  {Vinci}, \citenamefont {Chiaramello}, \citenamefont {Grassi}, \citenamefont
  {Fl{\'e}}, \citenamefont {Bouchard}, \citenamefont {Plotnikov}, \citenamefont
  {Aunai} \emph {et~al.}}]{derouillat2018smilei}%
  \BibitemOpen
  \bibfield  {author} {\bibinfo {author} {\bibfnamefont {J.}~\bibnamefont
  {Derouillat}}, \bibinfo {author} {\bibfnamefont {A.}~\bibnamefont {Beck}},
  \bibinfo {author} {\bibfnamefont {F.}~\bibnamefont {P{\'e}rez}}, \bibinfo
  {author} {\bibfnamefont {T.}~\bibnamefont {Vinci}}, \bibinfo {author}
  {\bibfnamefont {M.}~\bibnamefont {Chiaramello}}, \bibinfo {author}
  {\bibfnamefont {A.}~\bibnamefont {Grassi}}, \bibinfo {author} {\bibfnamefont
  {M.}~\bibnamefont {Fl{\'e}}}, \bibinfo {author} {\bibfnamefont
  {G.}~\bibnamefont {Bouchard}}, \bibinfo {author} {\bibfnamefont
  {I.}~\bibnamefont {Plotnikov}}, \bibinfo {author} {\bibfnamefont
  {N.}~\bibnamefont {Aunai}},  \emph {et~al.},\ }\href@noop {} {\bibfield
  {journal} {\bibinfo  {journal} {Computer Physics Communications}\ }\textbf
  {\bibinfo {volume} {222}},\ \bibinfo {pages} {351} (\bibinfo {year}
  {2018})}\BibitemShut {NoStop}%
\bibitem [{\citenamefont {Nuter}\ \emph {et~al.}(2011)\citenamefont {Nuter},
  \citenamefont {Gremillet}, \citenamefont {Lefebvre}, \citenamefont
  {L{\'e}vy}, \citenamefont {Ceccotti},\ and\ \citenamefont
  {Martin}}]{nuter2011field}%
  \BibitemOpen
  \bibfield  {author} {\bibinfo {author} {\bibfnamefont {R.}~\bibnamefont
  {Nuter}}, \bibinfo {author} {\bibfnamefont {L.}~\bibnamefont {Gremillet}},
  \bibinfo {author} {\bibfnamefont {E.}~\bibnamefont {Lefebvre}}, \bibinfo
  {author} {\bibfnamefont {A.}~\bibnamefont {L{\'e}vy}}, \bibinfo {author}
  {\bibfnamefont {T.}~\bibnamefont {Ceccotti}}, \ and\ \bibinfo {author}
  {\bibfnamefont {P.}~\bibnamefont {Martin}},\ }\href@noop {} {\bibfield
  {journal} {\bibinfo  {journal} {Physics of Plasmas}\ }\textbf {\bibinfo
  {volume} {18}},\ \bibinfo {pages} {033107} (\bibinfo {year}
  {2011})}\BibitemShut {NoStop}%
\bibitem [{\citenamefont {Arber}\ \emph {et~al.}(2015)\citenamefont {Arber},
  \citenamefont {Bennett}, \citenamefont {Brady}, \citenamefont
  {Lawrence-Douglas}, \citenamefont {Ramsay}, \citenamefont {Sircombe},
  \citenamefont {Gillies}, \citenamefont {Evans}, \citenamefont {Schmitz},
  \citenamefont {Bell} \emph {et~al.}}]{arber2015contemporary}%
  \BibitemOpen
  \bibfield  {author} {\bibinfo {author} {\bibfnamefont {T.}~\bibnamefont
  {Arber}}, \bibinfo {author} {\bibfnamefont {K.}~\bibnamefont {Bennett}},
  \bibinfo {author} {\bibfnamefont {C.}~\bibnamefont {Brady}}, \bibinfo
  {author} {\bibfnamefont {A.}~\bibnamefont {Lawrence-Douglas}}, \bibinfo
  {author} {\bibfnamefont {M.}~\bibnamefont {Ramsay}}, \bibinfo {author}
  {\bibfnamefont {N.}~\bibnamefont {Sircombe}}, \bibinfo {author}
  {\bibfnamefont {P.}~\bibnamefont {Gillies}}, \bibinfo {author} {\bibfnamefont
  {R.}~\bibnamefont {Evans}}, \bibinfo {author} {\bibfnamefont
  {H.}~\bibnamefont {Schmitz}}, \bibinfo {author} {\bibfnamefont
  {A.}~\bibnamefont {Bell}},  \emph {et~al.},\ }\href@noop {} {\bibfield
  {journal} {\bibinfo  {journal} {Plasma Physics and Controlled Fusion}\
  }\textbf {\bibinfo {volume} {57}},\ \bibinfo {pages} {113001} (\bibinfo
  {year} {2015})}\BibitemShut {NoStop}%
\bibitem [{\citenamefont {Massimo}\ \emph {et~al.}(2016)\citenamefont
  {Massimo}, \citenamefont {Atzeni},\ and\ \citenamefont
  {Marocchino}}]{massimo2016comparisons}%
  \BibitemOpen
  \bibfield  {author} {\bibinfo {author} {\bibfnamefont {F.}~\bibnamefont
  {Massimo}}, \bibinfo {author} {\bibfnamefont {S.}~\bibnamefont {Atzeni}}, \
  and\ \bibinfo {author} {\bibfnamefont {A.}~\bibnamefont {Marocchino}},\
  }\href@noop {} {\bibfield  {journal} {\bibinfo  {journal} {Journal of
  Computational Physics}\ }\textbf {\bibinfo {volume} {327}},\ \bibinfo {pages}
  {841} (\bibinfo {year} {2016})}\BibitemShut {NoStop}%
\bibitem [{\citenamefont {Benzi}\ \emph {et~al.}(1992)\citenamefont {Benzi},
  \citenamefont {Succi},\ and\ \citenamefont {Vergassola}}]{benzi1992lattice}%
  \BibitemOpen
  \bibfield  {author} {\bibinfo {author} {\bibfnamefont {R.}~\bibnamefont
  {Benzi}}, \bibinfo {author} {\bibfnamefont {S.}~\bibnamefont {Succi}}, \ and\
  \bibinfo {author} {\bibfnamefont {M.}~\bibnamefont {Vergassola}},\
  }\href@noop {} {\bibfield  {journal} {\bibinfo  {journal} {Physics Reports}\
  }\textbf {\bibinfo {volume} {222}},\ \bibinfo {pages} {145} (\bibinfo {year}
  {1992})}\BibitemShut {NoStop}%
\bibitem [{\citenamefont {Succi}(2018)}]{succi2018lattice}%
  \BibitemOpen
  \bibfield  {author} {\bibinfo {author} {\bibfnamefont {S.}~\bibnamefont
  {Succi}},\ }\href@noop {} {\emph {\bibinfo {title} {The lattice Boltzmann
  equation: for complex states of flowing matter}}}\ (\bibinfo  {publisher}
  {Oxford University Press},\ \bibinfo {year} {2018})\BibitemShut {NoStop}%
\bibitem [{\citenamefont {Kr{\"u}ger}\ \emph {et~al.}(2017)\citenamefont
  {Kr{\"u}ger}, \citenamefont {Kusumaatmaja}, \citenamefont {Kuzmin},
  \citenamefont {Shardt}, \citenamefont {Silva},\ and\ \citenamefont
  {Viggen}}]{kruger2017lattice}%
  \BibitemOpen
  \bibfield  {author} {\bibinfo {author} {\bibfnamefont {T.}~\bibnamefont
  {Kr{\"u}ger}}, \bibinfo {author} {\bibfnamefont {H.}~\bibnamefont
  {Kusumaatmaja}}, \bibinfo {author} {\bibfnamefont {A.}~\bibnamefont
  {Kuzmin}}, \bibinfo {author} {\bibfnamefont {O.}~\bibnamefont {Shardt}},
  \bibinfo {author} {\bibfnamefont {G.}~\bibnamefont {Silva}}, \ and\ \bibinfo
  {author} {\bibfnamefont {E.~M.}\ \bibnamefont {Viggen}},\ }\href@noop {}
  {\emph {\bibinfo {title} {The lattice Boltzmann method}}},\ Vol.~\bibinfo
  {volume} {10}\ (\bibinfo  {publisher} {Springer},\ \bibinfo {year}
  {2017})\BibitemShut {NoStop}%
\bibitem [{\citenamefont {Marocchino}\ \emph {et~al.}(2016)\citenamefont
  {Marocchino}, \citenamefont {Massimo}, \citenamefont {Rossi}, \citenamefont
  {Chiadroni},\ and\ \citenamefont {Ferrario}}]{marocchino2016efficient}%
  \BibitemOpen
  \bibfield  {author} {\bibinfo {author} {\bibfnamefont {A.}~\bibnamefont
  {Marocchino}}, \bibinfo {author} {\bibfnamefont {F.}~\bibnamefont {Massimo}},
  \bibinfo {author} {\bibfnamefont {A.}~\bibnamefont {Rossi}}, \bibinfo
  {author} {\bibfnamefont {E.}~\bibnamefont {Chiadroni}}, \ and\ \bibinfo
  {author} {\bibfnamefont {M.}~\bibnamefont {Ferrario}},\ }\href@noop {}
  {\bibfield  {journal} {\bibinfo  {journal} {Nuclear Instruments and Methods
  in Physics Research Section A: Accelerators, Spectrometers, Detectors and
  Associated Equipment}\ }\textbf {\bibinfo {volume} {829}},\ \bibinfo {pages}
  {386} (\bibinfo {year} {2016})}\BibitemShut {NoStop}%
\bibitem [{\citenamefont {Boris}\ and\ \citenamefont
  {Book}(1973)}]{boris1973flux}%
  \BibitemOpen
  \bibfield  {author} {\bibinfo {author} {\bibfnamefont {J.~P.}\ \bibnamefont
  {Boris}}\ and\ \bibinfo {author} {\bibfnamefont {D.~L.}\ \bibnamefont
  {Book}},\ }\href@noop {} {\bibfield  {journal} {\bibinfo  {journal} {Journal
  of computational physics}\ }\textbf {\bibinfo {volume} {11}},\ \bibinfo
  {pages} {38} (\bibinfo {year} {1973})}\BibitemShut {NoStop}%
\bibitem [{\citenamefont {Yee}(1966)}]{yee1966numerical}%
  \BibitemOpen
  \bibfield  {author} {\bibinfo {author} {\bibfnamefont {K.}~\bibnamefont
  {Yee}},\ }\href@noop {} {\bibfield  {journal} {\bibinfo  {journal} {IEEE
  Transactions on antennas and propagation}\ }\textbf {\bibinfo {volume}
  {14}},\ \bibinfo {pages} {302} (\bibinfo {year} {1966})}\BibitemShut
  {NoStop}%
\bibitem [{\citenamefont {Chen}\ and\ \citenamefont
  {Doolen}(1998)}]{chen1998lattice}%
  \BibitemOpen
  \bibfield  {author} {\bibinfo {author} {\bibfnamefont {S.}~\bibnamefont
  {Chen}}\ and\ \bibinfo {author} {\bibfnamefont {G.~D.}\ \bibnamefont
  {Doolen}},\ }\href@noop {} {\bibfield  {journal} {\bibinfo  {journal} {Annual
  review of fluid mechanics}\ }\textbf {\bibinfo {volume} {30}},\ \bibinfo
  {pages} {329} (\bibinfo {year} {1998})}\BibitemShut {NoStop}%
\bibitem [{\citenamefont {Bhatnagar}\ \emph {et~al.}(1954)\citenamefont
  {Bhatnagar}, \citenamefont {Gross},\ and\ \citenamefont
  {Krook}}]{bhatnagar1954model}%
  \BibitemOpen
  \bibfield  {author} {\bibinfo {author} {\bibfnamefont {P.~L.}\ \bibnamefont
  {Bhatnagar}}, \bibinfo {author} {\bibfnamefont {E.~P.}\ \bibnamefont
  {Gross}}, \ and\ \bibinfo {author} {\bibfnamefont {M.}~\bibnamefont
  {Krook}},\ }\href@noop {} {\bibfield  {journal} {\bibinfo  {journal}
  {Physical Review}\ }\textbf {\bibinfo {volume} {94}},\ \bibinfo {pages} {511}
  (\bibinfo {year} {1954})}\BibitemShut {NoStop}%
\bibitem [{\citenamefont {Srivastava}\ \emph {et~al.}(2013)\citenamefont
  {Srivastava}, \citenamefont {Perlekar}, \citenamefont {ten Thije~Boonkkamp},
  \citenamefont {Verma},\ and\ \citenamefont
  {Toschi}}]{srivastava2013axisymmetric}%
  \BibitemOpen
  \bibfield  {author} {\bibinfo {author} {\bibfnamefont {S.}~\bibnamefont
  {Srivastava}}, \bibinfo {author} {\bibfnamefont {P.}~\bibnamefont
  {Perlekar}}, \bibinfo {author} {\bibfnamefont {J.~H.}\ \bibnamefont {ten
  Thije~Boonkkamp}}, \bibinfo {author} {\bibfnamefont {N.}~\bibnamefont
  {Verma}}, \ and\ \bibinfo {author} {\bibfnamefont {F.}~\bibnamefont
  {Toschi}},\ }\href@noop {} {\bibfield  {journal} {\bibinfo  {journal}
  {Physical Review E}\ }\textbf {\bibinfo {volume} {88}},\ \bibinfo {pages}
  {013309} (\bibinfo {year} {2013})}\BibitemShut {NoStop}%
\bibitem [{\citenamefont {Premnath}\ and\ \citenamefont
  {Abraham}(2005)}]{premnath2005lattice}%
  \BibitemOpen
  \bibfield  {author} {\bibinfo {author} {\bibfnamefont {K.~N.}\ \bibnamefont
  {Premnath}}\ and\ \bibinfo {author} {\bibfnamefont {J.}~\bibnamefont
  {Abraham}},\ }\href@noop {} {\bibfield  {journal} {\bibinfo  {journal}
  {Physical Review E}\ }\textbf {\bibinfo {volume} {71}},\ \bibinfo {pages}
  {056706} (\bibinfo {year} {2005})}\BibitemShut {NoStop}%
\bibitem [{\citenamefont {Zhou}(2011)}]{zhou2011axisymmetric}%
  \BibitemOpen
  \bibfield  {author} {\bibinfo {author} {\bibfnamefont {J.~G.}\ \bibnamefont
  {Zhou}},\ }\href@noop {} {\bibfield  {journal} {\bibinfo  {journal} {Physical
  review E}\ }\textbf {\bibinfo {volume} {84}},\ \bibinfo {pages} {036704}
  (\bibinfo {year} {2011})}\BibitemShut {NoStop}%
\bibitem [{\citenamefont {Rosenzweig}(1987)}]{rosenzweig1987nonlinear}%
  \BibitemOpen
  \bibfield  {author} {\bibinfo {author} {\bibfnamefont {J.}~\bibnamefont
  {Rosenzweig}},\ }\href@noop {} {\bibfield  {journal} {\bibinfo  {journal}
  {IEEE transactions on plasma science}\ }\textbf {\bibinfo {volume} {15}},\
  \bibinfo {pages} {186} (\bibinfo {year} {1987})}\BibitemShut {NoStop}%
\bibitem [{\citenamefont {Schroeder}\ \emph {et~al.}(2005)\citenamefont
  {Schroeder}, \citenamefont {Esarey},\ and\ \citenamefont
  {Shadwick}}]{schroeder2005warm}%
  \BibitemOpen
  \bibfield  {author} {\bibinfo {author} {\bibfnamefont {C.}~\bibnamefont
  {Schroeder}}, \bibinfo {author} {\bibfnamefont {E.}~\bibnamefont {Esarey}}, \
  and\ \bibinfo {author} {\bibfnamefont {B.}~\bibnamefont {Shadwick}},\
  }\href@noop {} {\bibfield  {journal} {\bibinfo  {journal} {Physical Review
  E}\ }\textbf {\bibinfo {volume} {72}},\ \bibinfo {pages} {055401} (\bibinfo
  {year} {2005})}\BibitemShut {NoStop}%
\bibitem [{\citenamefont {Schroeder}\ and\ \citenamefont
  {Esarey}(2010)}]{schroeder2010relativistic}%
  \BibitemOpen
  \bibfield  {author} {\bibinfo {author} {\bibfnamefont {C.~B.}\ \bibnamefont
  {Schroeder}}\ and\ \bibinfo {author} {\bibfnamefont {E.}~\bibnamefont
  {Esarey}},\ }\href@noop {} {\bibfield  {journal} {\bibinfo  {journal}
  {Physical Review E}\ }\textbf {\bibinfo {volume} {81}},\ \bibinfo {pages}
  {056403} (\bibinfo {year} {2010})}\BibitemShut {NoStop}%
\bibitem [{\citenamefont {Zalesak}(1979)}]{zalesak1979fully}%
  \BibitemOpen
  \bibfield  {author} {\bibinfo {author} {\bibfnamefont {S.~T.}\ \bibnamefont
  {Zalesak}},\ }\href@noop {} {\bibfield  {journal} {\bibinfo  {journal}
  {Journal of computational physics}\ }\textbf {\bibinfo {volume} {31}},\
  \bibinfo {pages} {335} (\bibinfo {year} {1979})}\BibitemShut {NoStop}%
\bibitem [{\citenamefont {Mehrling}\ \emph {et~al.}(2018)\citenamefont
  {Mehrling}, \citenamefont {Benedetti}, \citenamefont {Schroeder},
  \citenamefont {Martinez De La~Ossa}, \citenamefont {Osterhoff}, \citenamefont
  {Esarey},\ and\ \citenamefont {Leemans}}]{mehrling2018accurate}%
  \BibitemOpen
  \bibfield  {author} {\bibinfo {author} {\bibfnamefont {T.}~\bibnamefont
  {Mehrling}}, \bibinfo {author} {\bibfnamefont {C.}~\bibnamefont {Benedetti}},
  \bibinfo {author} {\bibfnamefont {C.}~\bibnamefont {Schroeder}}, \bibinfo
  {author} {\bibfnamefont {A.}~\bibnamefont {Martinez De La~Ossa}}, \bibinfo
  {author} {\bibfnamefont {J.}~\bibnamefont {Osterhoff}}, \bibinfo {author}
  {\bibfnamefont {E.}~\bibnamefont {Esarey}}, \ and\ \bibinfo {author}
  {\bibfnamefont {W.}~\bibnamefont {Leemans}},\ }\href@noop {} {\bibfield
  {journal} {\bibinfo  {journal} {Physics of Plasmas}\ }\textbf {\bibinfo
  {volume} {25}},\ \bibinfo {pages} {056703} (\bibinfo {year}
  {2018})}\BibitemShut {NoStop}%
\bibitem [{\citenamefont {Adli}(2019)}]{adli2019plasma}%
  \BibitemOpen
  \bibfield  {author} {\bibinfo {author} {\bibfnamefont {E.}~\bibnamefont
  {Adli}},\ }\href@noop {} {\bibfield  {journal} {\bibinfo  {journal}
  {Philosophical Transactions of the Royal Society A}\ }\textbf {\bibinfo
  {volume} {377}},\ \bibinfo {pages} {20180419} (\bibinfo {year}
  {2019})}\BibitemShut {NoStop}%
\bibitem [{\citenamefont {Gabbana}\ \emph {et~al.}(2020)\citenamefont
  {Gabbana}, \citenamefont {Simeoni}, \citenamefont {Succi},\ and\
  \citenamefont {Tripiccione}}]{gabbana2020relativistic}%
  \BibitemOpen
  \bibfield  {author} {\bibinfo {author} {\bibfnamefont {A.}~\bibnamefont
  {Gabbana}}, \bibinfo {author} {\bibfnamefont {D.}~\bibnamefont {Simeoni}},
  \bibinfo {author} {\bibfnamefont {S.}~\bibnamefont {Succi}}, \ and\ \bibinfo
  {author} {\bibfnamefont {R.}~\bibnamefont {Tripiccione}},\ }\href@noop {}
  {\bibfield  {journal} {\bibinfo  {journal} {Physics Reports}\ }\textbf
  {\bibinfo {volume} {863}},\ \bibinfo {pages} {1} (\bibinfo {year}
  {2020})}\BibitemShut {NoStop}%
\bibitem [{\citenamefont {Yakimenko}\ \emph {et~al.}(2019)\citenamefont
  {Yakimenko}, \citenamefont {Alsberg}, \citenamefont {Bong}, \citenamefont
  {Bouchard}, \citenamefont {Clarke}, \citenamefont {Emma}, \citenamefont
  {Green}, \citenamefont {Hast}, \citenamefont {Hogan}, \citenamefont {Seabury}
  \emph {et~al.}}]{yakimenko2019facet}%
  \BibitemOpen
  \bibfield  {author} {\bibinfo {author} {\bibfnamefont {V.}~\bibnamefont
  {Yakimenko}}, \bibinfo {author} {\bibfnamefont {L.}~\bibnamefont {Alsberg}},
  \bibinfo {author} {\bibfnamefont {E.}~\bibnamefont {Bong}}, \bibinfo {author}
  {\bibfnamefont {G.}~\bibnamefont {Bouchard}}, \bibinfo {author}
  {\bibfnamefont {C.}~\bibnamefont {Clarke}}, \bibinfo {author} {\bibfnamefont
  {C.}~\bibnamefont {Emma}}, \bibinfo {author} {\bibfnamefont {S.}~\bibnamefont
  {Green}}, \bibinfo {author} {\bibfnamefont {C.}~\bibnamefont {Hast}},
  \bibinfo {author} {\bibfnamefont {M.}~\bibnamefont {Hogan}}, \bibinfo
  {author} {\bibfnamefont {J.}~\bibnamefont {Seabury}},  \emph {et~al.},\
  }\href@noop {} {\bibfield  {journal} {\bibinfo  {journal} {Physical Review
  Accelerators and Beams}\ }\textbf {\bibinfo {volume} {22}},\ \bibinfo {pages}
  {101301} (\bibinfo {year} {2019})}\BibitemShut {NoStop}%
\bibitem [{\citenamefont {Assmann}\ \emph {et~al.}(2020)\citenamefont
  {Assmann}, \citenamefont {Weikum}, \citenamefont {Akhter}, \citenamefont
  {Alesini}, \citenamefont {Alexandrova}, \citenamefont {Anania}, \citenamefont
  {Andreev}, \citenamefont {Andriyash}, \citenamefont {Artioli}, \citenamefont
  {Aschikhin} \emph {et~al.}}]{assmann2020eupraxia}%
  \BibitemOpen
  \bibfield  {author} {\bibinfo {author} {\bibfnamefont {R.}~\bibnamefont
  {Assmann}}, \bibinfo {author} {\bibfnamefont {M.}~\bibnamefont {Weikum}},
  \bibinfo {author} {\bibfnamefont {T.}~\bibnamefont {Akhter}}, \bibinfo
  {author} {\bibfnamefont {D.}~\bibnamefont {Alesini}}, \bibinfo {author}
  {\bibfnamefont {A.}~\bibnamefont {Alexandrova}}, \bibinfo {author}
  {\bibfnamefont {M.}~\bibnamefont {Anania}}, \bibinfo {author} {\bibfnamefont
  {N.}~\bibnamefont {Andreev}}, \bibinfo {author} {\bibfnamefont
  {I.}~\bibnamefont {Andriyash}}, \bibinfo {author} {\bibfnamefont
  {M.}~\bibnamefont {Artioli}}, \bibinfo {author} {\bibfnamefont
  {A.}~\bibnamefont {Aschikhin}},  \emph {et~al.},\ }\href@noop {} {\bibfield
  {journal} {\bibinfo  {journal} {The European Physical Journal Special
  Topics}\ }\textbf {\bibinfo {volume} {229}},\ \bibinfo {pages} {3675}
  (\bibinfo {year} {2020})}\BibitemShut {NoStop}%
\bibitem [{\citenamefont {Aschikhin}\ \emph {et~al.}(2016)\citenamefont
  {Aschikhin}, \citenamefont {Behrens}, \citenamefont {Bohlen}, \citenamefont
  {Dale}, \citenamefont {Delbos}, \citenamefont {Di~Lucchio}, \citenamefont
  {Elsen}, \citenamefont {Erbe}, \citenamefont {Felber}, \citenamefont {Foster}
  \emph {et~al.}}]{aschikhin2016flashforward}%
  \BibitemOpen
  \bibfield  {author} {\bibinfo {author} {\bibfnamefont {A.}~\bibnamefont
  {Aschikhin}}, \bibinfo {author} {\bibfnamefont {C.}~\bibnamefont {Behrens}},
  \bibinfo {author} {\bibfnamefont {S.}~\bibnamefont {Bohlen}}, \bibinfo
  {author} {\bibfnamefont {J.}~\bibnamefont {Dale}}, \bibinfo {author}
  {\bibfnamefont {N.}~\bibnamefont {Delbos}}, \bibinfo {author} {\bibfnamefont
  {L.}~\bibnamefont {Di~Lucchio}}, \bibinfo {author} {\bibfnamefont
  {E.}~\bibnamefont {Elsen}}, \bibinfo {author} {\bibfnamefont {J.-H.}\
  \bibnamefont {Erbe}}, \bibinfo {author} {\bibfnamefont {M.}~\bibnamefont
  {Felber}}, \bibinfo {author} {\bibfnamefont {B.}~\bibnamefont {Foster}},
  \emph {et~al.},\ }\href@noop {} {\bibfield  {journal} {\bibinfo  {journal}
  {Nuclear Instruments and Methods in Physics Research Section A: Accelerators,
  Spectrometers, Detectors and Associated Equipment}\ }\textbf {\bibinfo
  {volume} {806}},\ \bibinfo {pages} {175} (\bibinfo {year}
  {2016})}\BibitemShut {NoStop}%
\end{thebibliography}%

\end{document}